# Autonomous Control of a Quadrotor-Manipulator

*Application of Extended State Disturbance Observer*

Abdulmajeed M. Kabir
kbmajeed1@gmail.com,
Department of Systems Engineering,
King Fahd University of Petroleum and Minerals

*Abstract*—In this work, the autonomous control of a quadrotor-manipulator unmanned aerial vehicle is treated using an extended dynamic model. Due to persistent aerodynamic disturbances and dynamic couplings, the control of a quadrotor-manipulator unit is cumbersome. Active disturbance rejection control (ADRC) is proposed for the cancellation of the undesirable dynamics while a PD controller is developed for set-point tracking. Parameters were tuned using Gradient Descent-Sequential Quadratic Programming in Simulink design optimization.

## I. INTRODUCTION

Unmanned Aerial Vehicle (UAV) can simply be defined as a space-traversing vehicle that flies without a human pilot or crew and can be remotely controlled or can fly autonomously. There has been a sharp rise in the development, control and applications of quadrotors in recent years and the field of aerial robotics is fast-growing with engineers and hobbyists actively involved. A broad study on UAVs can be found in [1]. Quadrotors are small-sized, simple and highly maneuverable vehicles operating in three dimensional space, capable of translations $x$, $y$, $z$ and rotations $\phi$, $\theta$, and $\psi$ resulting in six degrees of freedom. It is worthy of note that with four rotors as input actuators and six degrees of freedom, the quadrotor is an under-actuated system and as a result, there exists constraints in maneuvering. The translation and orientation of the quadrotor is controlled by intermittent increase and or decrease in thrusts on the four motor actuators via the electronic speed controllers (ESCs).

Quadrotor-manipulator (QM) systems on the other hand are quadrotor systems equipped with a robotic arm. The addition of a robotic arm induces coupled dynamics and other effects which makes control more complicated. The main intrinsic effects of an added robotic manipulator to the existing quadrotor is mainly: (a) a change in the center of mass, (b) change in inertia properties and (c) change in overall mass. Apart from these, extrinsic influences such as (c) aerodynamic drag, (d) ground effect and (e) wind disturbances influence the quadrotor dynamics negatively. Appropriate control of quadrotor-manipulator systems is important as they are applied in several delicate applications such as: load transportation, disaster response, autonomous manipulation, aerial photography, surveillance, aerial inspection, animal tracking, and even in the nuclear industry.

A number of authors have solved the quadrotor-manipulator control problem. In [2], [3] the authors proposed a PI-D algorithm for the dexterous control of a mobile manipulating quadrotor while [4] designed a passivity based controller for a planar quadrotor-manipulator system. In [5], the authors developed a mass estimator and consequently developed a controller using Lyapunov theory. [6] utilized model predictive control, [7] combined sliding mode control with $H_\infty$ for the control of a rotorcraft aerial vehicle while [8] presented a compliant control scheme for a quadrotor helicopter equipped with an n-DOF manipulator. In [9], the quadrotor and manipulator dynamics were handled separately with the use of adaptive and PD control while the manipulator was controlled using PID controller. Similarly, [10] implemented a hybrid model predictive control algorithm for the control of an unmanned rotorcraft interacting with the environment. The work by [11] involved the use of variable parameter integral backstepping (VPIB) for an outdoor aerial manipulator. [12] presented estimation techniques for the mass properties of an aerial grasping and manipulator system with a unique manipulator and a controller based on the estimated mass parameters. Furthermore, interesting reports on helicopter stability under payload can be found in [13], [14]. As a result, only rare published works exists on dedicated control algorithms that compensates for all undesirable intrinsic and extrinsic dynamic effects of a quadrotor-manipulator system. In this work, the ADRC controller is presented to estimate and compensate for the total disturbance of the quadrotor system with a 1-DOF prismatic manipulator. This is based on a disturbance observer technique that has not been applied previously to the quadrotor manipulator model presented in Section II, (11).

Disturbance observers are known for their use in the estimation of unmeasurable dynamics in a plant. An ADRC controller [15] as the name implies actively cancels disturbances and unwanted dynamics. It is based on the concept of an extended state observer (ESO). An extended state observer is a type of disturbance observer that estimates combined or total disturbances in a plant by introducing a fictitious state (also known as the extended state) that represents the unknown or unwanted dynamics. After estimation of the unknown dynamics, a feedback linearization-type operation is used to cancel out the total disturbance from the plant dynamics. This implies that the control affine plant transforms to a linear feedback integrator system. As a result, linear control theory can be applied. It is thus intuitive that the performance of the controller is dependent on the accuracy, speed and performance of the ESO.

This paper is organized as follows: Section II formulates the problem, mentions the undesirable effects that needs to be estimated, compensated and cancelled. The extended dynamic model of the quadrotor-manipulator system is presented and the control objective is stated. Section III describes the formulation of the ADRC controller. Section IV presents the application of the techniques to the problem and parameter optimization. Section V describes the results obtained from Matlab simulation while Section VI contains the summary of results obtained and the conclusion of the research work.

## II. PROBLEM FORMULATION

Controlling an autonomous quadrotor-manipulator is a delicate process and is best handled by a control architecture that is robust to bounded parameter changes and external disturbances. Quadrotor-manipulator systems are influenced by intrinsic and extrinsic dynamic disturbances some of which are detailed in the next section. The main intrinsic disturbances considered in this work include: change in center of mass (CoM effect), change in moment of inertia and change in overall mass. On the other hand, the extrinsic disturbances considered are: ground wake effect, aerodynamic drag and wind. Finally, the ADRC controller observes, estimates and actively compensates and eliminates these effects from the dynamics of the quadrotor-manipulator (QM) for effective control. The ADRC parameters also need to be effectively tuned. The disturbance models are presented next and thereafter; a complete model is finally obtained.

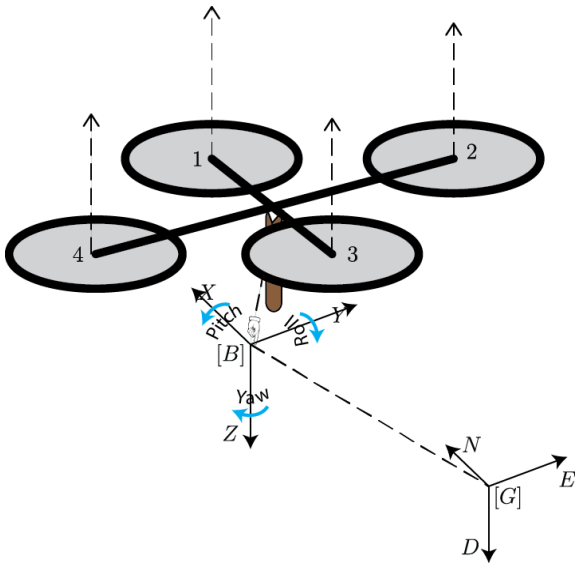

Fig. 1. *Quadrotor-Manipulator System* .

For the sake of clarity, the quadrotor's state vectors are predefined. $\phi$ is the roll angle, $\theta$ is the pitch angle, $\psi$ is the yaw angle, $z$ is the altitude, $x, y$ are the quadrotor translation. $\phi = x_1, \theta = x_3, \psi = x_5, z = x_7, x = x_9, y = x_{11}$. The derivatives of the states are velocities $\dot{\phi}, \dot{\theta}, \dot{\psi}, \dot{z}, \dot{x}, \dot{y} = x_2, x_4, x_6, x_8, x_{10}, x_{12}$ and similarly; $\dot{x}_1 = x_2, \dot{x}_3 = x_4, \dot{x}_5 = x_6, \dot{x}_7 = x_8, \dot{x}_9 = x_{10}$ and $\dot{x}_{11} = x_{12}$.

### A. Aerodynamic Drag

Authors in [16]–[19] have discussed the aerodynamic effects on UAV vehicles. In [17], blade flapping, induced drag, translational drag and parasitic drag were highlighted as aerodynamic forces to be considered while [16] studied the effects of external perturbations such as drag force and wind disturbance on the flight performance of a quadrotor model and thereafter implemented inverse dynamics and linear PID controller. Aerodynamic drag ($D_i$) is a resisting force preventing motion of a vehicle interacting with surrounding fluid. The magnitude of aerodynamic drag is proportional to the velocity of the QM and the constant of proportionality is the

$$D_i = k_i x_i, \qquad i = 2,4,6,8,10,12 \qquad (1)$$

drag coefficient ($k_i$). The model of drag used as in [20], is given by (1) with $m$ standing for mass and $x_i$ represents velocity.

### B. Ground Wake Effect

Ground effect [21] is an important aerodynamic effect that needs to be taken into consideration when landing quadrotor vehicles. As highlighted in [13], ground effect is a phenomenon wherein the *wake* of a helicopter rotor pushes a cushion of air resisting the helicopter's descent, thereby, creating a form of damping or cushion effect. This makes smooth descent of helicopter aircrafts difficult to achieve. Failure to compensate for this effect will cause plunging or bouncing (Figure 5) off the helicopter's desired landing target [22].

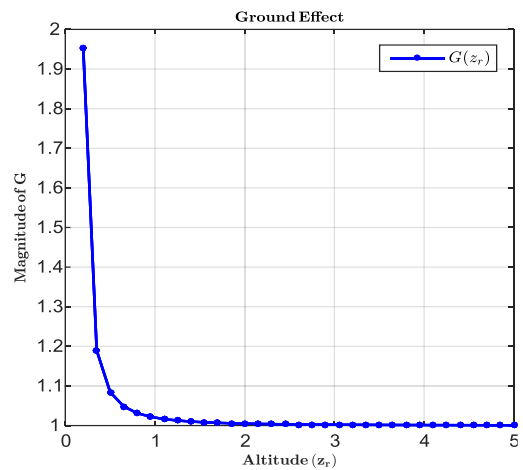

Fig. 2. *Quadrotor-Manipulator System* .

The authors in [23] presented the relationship between quadrotor vertical thrust in and out of the ground effect zone (2).

$$U_1 = U_z\left(1 - \rho\left(\frac{r}{4z_r}\right)^2\right) \qquad (2)$$

$\rho = 8.6$ was determined experimentally with a quadrotor with rotor radius $r = 0.1905$m. $U_z$ is now regarded as a virtual control input to the altitude subsystem, $U_1$ is the actual control input, $z_r = z$ is the vertical distance from rotor to ground (altitude).

A new variable $G$ is introduced (3) which is a scaling factor for the altitude thrust.

$$G = \frac{1}{\left(1 - \rho \left(\frac{r}{4z_r}\right)^2\right)} \quad (3)$$

$G$ comes to live at a particularly low altitude determined by $r$ and $z_r$. Above this range, $G = 1$ as shown in Figure 2. The effect of $G$ can be observed in Figure 2 as influential at a particular range of values for $z < 5m$. $z_r = 0$ is avoided due to presence of landing skids which constrains $z_r > 0$. It was experimentally proven in [23] that the ground effect is prominent at altitudes below 4m.

### C. Wind Disturbance

In [24], the authors discussed the estimation of wind effects on quadrotor using wind tunnel tests. From the paper [25], it was mentioned that wind effect on quadrotor flight control can be significant and lead to instabilities, thereafter, a Dryden wind-gust model was presented. In order to improve the position control of the quadrotor, wind effects can be modelled approximately and compensated in the controller design. The Dryden wind-gust model - a summation of sinusoidal excitations, is commonly used in this regard. In this work however, the wind disturbance is modelled using a simple sinusoidal disturbance model (4) as presented in [26]:

$$\varpi_i(t) = \alpha_i + \beta_i \sin(nt) \quad (4)$$

$\varpi_i(t)$ is a time dependent estimate of the wind vector at time $t$. $a_i$ and $b_i$ represent the disturbance magnitude.

### D. Total Mass and Center of Mass Effect

The total mass $(m)$ of the quadrotor-manipulator system is the sum of the quadrotor $(m_q)$ and the manipulator masses $(m_r)$.

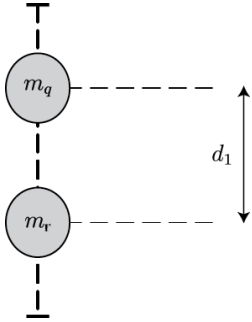

Fig. 3. *Mass Center*.

Making sure the quadrotor and prismatic arm are vertically aligned on the same z-axis, the resulting center of mass $(x_G = 0, y_G = 0, z_G)$ of the quadrotor-manipulator (Figure 3) can be calculated using the Formula (5):

$$z_G = \frac{m_q d_0 + m_r d_1}{m_q + m_r} \quad (5)$$

$m_q$, is the quadrotor mass, $m_r$, is the mass of the robotic arm, $d_0 = 0$, is taken as the reference point at the quadrotor's center of mass. $d_1$ is the distance of the robotic arm's center of mass from $d_0$.

The effect of the added manipulator and payload on the center of mass (CoM) of the quadrotor was studied in [27] and [28]. It was mentioned that the shift in CoM introduces additional accelerations and velocities. Although mass trimming can be done to balance out the CoM, this procedure is very tedious. Also, if sensors are not exactly aligned to the CoM, they would provide erroneous data. The effect of the added manipulator and load as described in [27] is the introduction of additional dynamics. The induced dynamics in this case is caused by the CoM effect along the $z$-axis and the resulting CoM effect equations along the translation and orientation dynamics are given in (6). $z_G$ is the shift in CoM along the $z$-axis.

$$\begin{aligned}
\mho_{\phi \Leftrightarrow x_2} &= -mz_G(\dot{x}_{12} + x_{10}x_6) \\
\mho_{\theta \Leftrightarrow x_4} &= mz_G(\dot{x}_{10} - x_{12}x_6) \\
\mho_{\psi \Leftrightarrow x_6} &= mz_G(x_{12}x_4 + x_4x_2) \\
\mho_{z \Leftrightarrow x_8} &= z_G(x_2^2 - x_4^2) \\
\mho_{x \Leftrightarrow x_{10}} &= -z_G(x_2x_6 - \dot{x}_4) \\
\mho_{y \Leftrightarrow x_{12}} &= -z_G(x_4x_6 - \dot{x}_2)
\end{aligned} \quad (6)$$

### E. Moment of Inertia Effect

The moment of inertia for the QM system can be calculated by considering a cross axis cylindrical body as the quadrotor and a cuboid as the prismatic robotic arm.

The moment of inertia of the quadrotor cross-axis frame is represented as two cylinders $(q)$ with the following moment of inertia properties, dimensions and mass

$$\begin{aligned}
I_{xq} &= m_q \left(\frac{R_q^2}{4} + \frac{L_q^2}{12} + \frac{R_q^2}{2}\right) \\
I_{yq} &= m_q \left(\frac{R_q^2}{4} + \frac{L_q^2}{12} + \frac{R_q^2}{2}\right) \\
I_{zq} &= m_q \left(\frac{R_q^2}{4} + \frac{L_q^2}{12} + \frac{R_q^2}{4} + \frac{L_q^2}{12}\right)
\end{aligned} \quad (7)$$

The moment of inertia of the robotic arm $(r)$ is represented as a cuboid with the following moment of inertia properties:

$$\begin{aligned}
I_{xr} &= m_r \left(\frac{W_r^2}{12} + \frac{H_r^2}{12} + D_r^2\right) \\
I_{yr} &= m_r \left(\frac{L_r^2}{12} + \frac{H_r^2}{12} + D_r^2\right) \\
I_{zr} &= m_r \left(\frac{L_r^2}{2} + \frac{W_r^2}{2}\right)
\end{aligned} \quad (8)$$

$R_i$-radius, $L_i$-length, $W_i$-width, $H_i$-height, $D_i$-distance to the quadrotor's central axis are the geometric dimensions of the quadrotor-manipulator parts.

The new moment of inertia is now obtained by adding the individual quadrotor and robot arm inertias as proposed by parallel-axis theorem.

$$\begin{aligned} I_{xx} &= I_{xq} + I_{xr} \\ I_{yy} &= I_{yq} + I_{yr} \\ I_{zz} &= I_{zq} + I_{zr} \end{aligned} \quad (9)$$

Thus, the resulting change in the moment of inertia (combined moment of inertia) while keeping the assumption of symmetry true can be described by the Inertia matrix $J$ is given in (10):

$$J = \begin{pmatrix} I_{xx} & 0 & 0 \\ 0 & I_{yy} & 0 \\ 0 & 0 & I_{zz} \end{pmatrix} \quad (10)$$

*F. Dynamic Model Equations of a Quadrotor-Manipulator*

The final extended dynamic model of the QM in state space is thus (11):

$$\begin{bmatrix} \dot{x}_1 \\ \dot{x}_2 \\ \dot{x}_3 \\ \dot{x}_4 \\ \dot{x}_5 \\ \dot{x}_6 \\ \dot{x}_7 \\ \dot{x}_8 \\ \dot{x}_9 \\ \dot{x}_{10} \\ \dot{x}_{11} \\ \dot{x}_{12} \end{bmatrix} = \begin{bmatrix} x_2 \\ U_2 a_6 - a_2 x_4 \Omega_r + a_1 x_4 x_6 + \Delta_a \\ x_4 \\ U_3 a_7 + a_4 x_2 \Omega_r + a_3 x_2 x_6 + \Delta_b \\ x_6 \\ U_4 a_8 + a_5 x_2 x_4 + \Delta_c \\ x_8 \\ g - G \dfrac{U_1}{m}(\cos x_3 \, \cos x_1) + \Delta_d \\ x_{10} \\ -\dfrac{U_1}{m}(\sin x_1 \sin x_5 + \sin x_3 \cos x_1 \cos x_5) + \Delta_e \\ x_{12} \\ -\dfrac{U_1}{m}(\cos x_1 \sin x_3 \sin x_5 - \sin x_1 \cos x_5) + \Delta_f \end{bmatrix} \quad (11)$$

$$\begin{aligned} \Delta_a &= -\mho_{x_2} - k_{x_2} x_2 + \varpi_{x_2} \\ \Delta_b &= -\mho_{x_4} - k_{x_4} x_4 + \varpi_{x_4} \\ \Delta_c &= +\mho_{x_6} - k_{x_6} x_6 + \varpi_{x_6} \\ \Delta_d &= -\mho_{x_8} + k_{x_8} x_8 + \varpi_{x_8} \\ \Delta_e &= -\mho_{x_{10}} - k_{x_{10}} x_{10} + \varpi_{x_{10}} \\ \Delta_f &= -\mho_{x_{12}} - k_{x_{12}} x_{12} + \varpi_{x_{12}} \end{aligned} \quad (12)$$

The corresponding state vector for representing $[\phi \; \dot{\phi} \; \theta \; \dot{\theta} \; \psi \; \dot{\psi} \; z \; \dot{z} \; x \; \dot{x} \; y \; \dot{y}]^T$ in state space is given by $[x_1 \; x_2 \; x_3 \; x_4 \; x_5 \; x_6 \; x_7 \; x_8 \; x_9 \; x_{10} \; x_{11} \; x_{12}]^T$. Where $x_1$ is the $\phi$-roll angle, $x_3$ is the $\theta$-pitch angle, $x_5$ is the $\psi$-yaw angle, $x_7$ is the $z$-altitude state, $x_9$ is the $x$-translation and $x_{11}$ is the y-translation. Similarly, $a_i$, contains the inertia model parameters: $a_1 = \dfrac{I_{yy} - I_{zz}}{I_{xx}}$, $a_2 = \dfrac{J_r}{I_{xx}}$, $a_3 = \dfrac{I_{zz} - I_{xx}}{I_{yy}}$, $a_4 \dfrac{J_r}{I_{yy}}$, $a_5 = \dfrac{I_{xx} - I_{yy}}{I_{zz}}$, $a_6 = \dfrac{l}{I_{xx}}$, $a_7 = \dfrac{l}{I_{yy}}$, $a_8 = \dfrac{1}{I_{zz}}$. $m$ is the mass of the QM. $\Omega_r = -\Omega 1 + \Omega 2 - \Omega 3 + \Omega 4$ is the relative rotor speeds characterized by (13):

$$\begin{bmatrix} U_1 \\ U_2 \\ U_3 \\ U_4 \end{bmatrix} = \begin{bmatrix} k_f & k_f & k_f & k_f \\ 0 & -k_f & 0 & k_f \\ k_f & 0 & -k_f & 0 \\ k_m & -k_m & k_m & -k_m \end{bmatrix} \begin{bmatrix} \Omega_1^2 \\ \Omega_2^2 \\ \Omega_3^2 \\ \Omega_4^2 \end{bmatrix} \quad (13)$$

$U_1$ is the altitude control signal, $U_2$ is responsible for the roll and $U_3$ for pitch while $U_4$ is the required control signal for the desired yaw (heading) angles. $k_f$ and $k_m$ are constants of aerodynamic force and moments respectively. Similarly, $G$ is the ground effect and $\Delta_{a \to f}$ encapsulates undesirable disturbances.

### III. ADRC CONTROLLER

Active disturbance rejection control (ADRC) is a relatively new control technique that has gained attention of the control community. The novel ADRC technique developed by Jinqing Han in [29]. ADRC control technique is based on a unique type of disturbance observer known as extended state observer (ESO). State observers, also known as estimators, are very useful in the design of modern control systems. They are used to estimate the internal variables of a dynamic plant using input and output signals only. Observers are used in flux estimation of A.C. induction motors. An interesting discussion on disturbance observers can be found in [30]. In [31], the ESO was classified as an efficient observer requiring the least amount of plant information for its operation compared to other disturbance observers such as the unknown input observer (UIO).

Obtaining accurate mathematical models for highly complex and nonlinear systems is usually a challenge. This is partly because dynamics could be coupled, nonlinear and or even stochastic. More so, the attenuation, compensation and elimination of disturbances from physical system is usually a key criterion in control design. The main operation of the ADRC is to estimate (using the ESO) and compensate for the effects of unknown dynamics and disturbances. This transforms a system in control-affine form (8) to a simple linear-feedback double-integrator. Stability analysis of the ADRC controller was studied by Qing Zheng in [31] and Wankun Zhou et al. in [32]. The analytical results obtained verifies the stability of the ADRC system which enhances confidence for its real-time implementation. Consequently, the ADRC has been applied in robot motion control [33], industrial heater [34], boiler units [35], electrical voltage regulation [36], marine steam turbine [37] and autonomous aerial vehicles [38].

## A. ADRC Controller Formulation

The most important system in describing the ADRC is the ESO. The formulation of the ADRC as obtained in [15], [31] and [33] is treated in this section. Given a second order single-input-single-output (SISO) system affine in control:

$$\ddot{x} = f(x_1, x_2, \omega(t), t) + bu$$

In state space form:

$$\begin{aligned} \dot{x}_1 &= x_2 \\ \dot{x}_2 &= f(x_1, x_2, \omega(t), t) + bu \\ y &= x_1 \end{aligned} \quad (14)$$

where $y \in \mathbb{R}$ is the plant output, measurable and to be controlled, $u \in \mathbb{R}$ is the input, and $f(x_1, x_2, \omega(t), t) = F(t)$ is a function of the plant's states: $x_i \in \mathbb{R}$, external disturbances $\omega$, and time $t$. $F(t)$ is regarded to as the total disturbance and assumed to be differentiable. The goal is usually to make $y$ track a desired signal or reference by manipulating $u$. Taking $F(t)$ as an additional state variable $x_3 = F(t)$ and denoting $\dot{F}(t) = \bar{G}(t)$, with $\bar{G}(t)$ unknown, the original plant in (14) is now described in extended or augmented form as:

$$\begin{aligned} \dot{x}_1 &= x_2 \\ \dot{x}_2 &= x_3 + bu \\ \dot{x}_3 &= \bar{G}(t) \\ y &= x_1 \end{aligned} \quad (15)$$

The ESO with linear gain parameters is also presented as:

$$\begin{aligned} \dot{\hat{x}}_1 &= \hat{x}_2 + p_1(x_1 - \hat{x}_1) \\ \dot{\hat{x}}_2 &= \hat{x}_3 + p_2(x_1 - \hat{x}_1) + \hat{b}u \\ \dot{\hat{x}}_3 &= \quad\quad p_3(x_1 - \hat{x}_1) \end{aligned} \quad (16)$$

$\hat{x}_1$ is the estimate of $x_1$, $\hat{x}_2$ is the estimate of $x_2$, $\hat{x}_3$ is the estimate of $F(t)$. $p_{i:1,2,3}$ are the observer gains (bandwidth) to be tuned. The observer gains can be tuned manually such that the characteristic polynomial $s^3 + p_1 s^2 + p_2 s + p_3$ is Hurwitz. As expected, larger observer gains result in more accurate state estimations. This however comes at the detriment of increased noise sensitivity. Thus, the appropriate observer bandwidth should be selected as a compromise between the tracking performance and noise tolerance. In essence, the stability margins, performance characteristics and noise sensitivity of the system can be put in check by an optimization approach to tuning the gains. Note that the ground effect factor $G$ is not the same as the extended state derivative $\bar{G}$.

The architecture of the ADRC controller is briefly depicted in Figure 3.

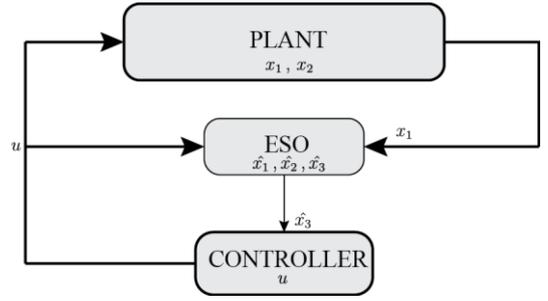

Fig. 4. Architecture for the ADRC Controller.

Using the control structure (17) in (14) eliminates the total

$$u = \frac{u_0 - F(t)}{\hat{b}} \quad (17)$$

disturbance from the plant (14) thereby reducing the plant to a simple linear feedback double-integrator (18).

$$\begin{aligned} \dot{x}_1 &= x_2 \\ \dot{x}_2 &= u_0 \\ \ddot{x} &= u_0, y = x_1, \boldsymbol{x} = [x_1, x_2]^{\mathrm{T}} \end{aligned} \quad (18)$$

It is now possible to control the system (14) via (17) using linear control theory for $u_0$. In this work, the PD controller was employed at this stage. The closed loop system is therefore:

$$\ddot{x} = (f(x_1, x_2, \omega(t), t) - F(t)) + u_0 \quad (19)$$

For a properly designed and tuned ESO,

$$f(x_1, x_2, \omega(t), t) - F(t) \approx 0. \quad (20)$$

The closer the left hand side of (16) is to zero, the closer the plant (14) is to an ideal double integrator (18). This closeness can be enforced by optimizing the ESO gains $p_{i:1,2,3}$.

The PD controller for the double integrator is given in (21):

$$u_0 = k_p(x_{1\mathrm{r}} - x_1) + k_d(\dot{x}_{1\mathrm{r}} - x_2) \quad (21)$$

## IV. QUADROTOR CONTROL

### A. Applying the ADRC for Landing

Now that the dedicated ADRC controller has been developed, it is applied to the QM problem. The QM dynamics is divided into four subsystems – roll ($\phi$), pitch ($\theta$), yaw ($\psi$) and altitude ($z$) subsystems. These subsystems capture the necessary dynamics for full control of the quadrotor. An ADRC formulation is generated for each of the subsystems. The $\hat{b}$ parameter represents a control signal perturbation and is crucial in the stability of the closed-loop system. For the subsystems,

$\hat{b}_\phi = a_6$, $\hat{b}_\theta = a_7$, $\hat{b}_\psi = a_8$, $\hat{b}_z = \frac{-G}{m}(\cos x_3 \cos x_1)$. The double integrator equivalent of each subsystem is controlled using Proportional-Derivative (PD) controller (21) which generates $u_0$ for the desired performance of (11). The use of a simple PD controller guarantees the system stability it also introduces two extra parameters $k_{1i}$ and $k_{2i}$ for tuning. Overall, there are three ESO parameters and two PD parameters per subsystem. Finally, a total of eleven parameters need tuning. To ease the computational requirements, the same ESO gains $(p_{1,2,3})$ are used for all subsystems while initializing the optimization task. Cost savings can be achieved by using estimated states instead of sensor readings, thus, higher order systems can be controlled using fewer sensors. This is another advantage of the ADRC controller.

### B. Parameter optimization

Parameter optimization is very crucial for any tuning problem. Numerous algorithms have been developed for optimization and utilized in a cornucopia of applications; even outside the control field. In [39], the particle swarm algorithm was applied for the optimization of mobile robot controller while [40] applied genetic algorithms to tune PID controllers. Optimization can be applied online (gains are dynamically fitted during process loop) or offline (static gains obtained are used in the process loop). The objective function is set to minimize or maximize a cost function. The use of evolutionary algorithms to optimize system parameters is widespread and often requires a tedious development of code to implement the algorithms.

A powerful and usually overlooked tool for parameter optimization; '*Response Optimization*' lies in Matlab (*v.2015b*) itself. The Simulink Design Optimization (SDO) software [41] automatically converts design requirements to a constrained optimization problem and then applies optimization techniques. The Simulink model is then iteratively simulated until the design requirements are satisfied. This SDO tool also has a graphical user interface where users can utilize click-programming to set decision variables and signal bounds. It can also be implemented in code. The signal bounds (optimization criteria) can be seen as the cost function and constraints of the problem. They are defined as piecewise linear bounds. Once all is set, the user chooses an appropriate solver such as: gradient descent, pattern search or simplex search. In order to speed up optimization, the parallel computing option can be activated. This tool is powerful and easy to use. The entire optimization task is very flexible.

In this work, Gradient Descent-Sequential Quadratic Programming combined with parallel computing were used to optimize the ESO and PD parameters such that the error in estimation is minimized and the quadrotor tracks a reference set-point appropriately. The results obtained using an Intel-Core i7 2.6 GHz CPU are summarized in Table 2.

## V. SIMULATION AND RESULTS

### A. Simulation criteria and cases

The QM manipulator parameters are given in Table 1 and the optimized controller simulation parameters are given in Table 2.

TABLE 1. QUADROTOR MANIPULATOR PARAMETERS

| No. | Quadrotor Parameters | | |
|---|---|---|---|
| | *Parameter* | *Symbol* | *Value* |
| 1 | Mass | $m$(kg) | 2 |
| 2 | Rotor radius | $r$(m) | 0.1905 |
| 3 | Aerodynamic drag coefficient | $k_{x2,4\ldots12}$ | 0.3729 |
| 4 | Ground effect coefficient | $\rho$ | 8.6 |
| 5 | Gravitational constant | $g(\text{ms}^{-2})$ | 9.81 |
| 6 | Inertia Parameters | $I_{xx}, I_{yy}$ | 0.018 |
| 7 | Inertia Parameters | $I_{zz}$ | 0.035 |
| 8 | Length of cross-axis | $l$(m) | 0.45 |
| 9 | Rotor Inertia | $J_r$ | $6 \times 10^{-3}$ |
| 10 | CoM shift along z-axis | $z_G$ | 0.08 |
| 11 | Wind disturbance parameters | $\alpha_i, \beta_i$ | 0.1, 1 |

TABLE 2. OPTIMIZED PARAMETERS

| No. | System Parameters | | |
|---|---|---|---|
| | *Type* | *Parameter* | *Value* |
| 1 | ESO Observer gain | $p_1$ | 29.5659 |
| 2 | ESO Observer gain | $p_2$ | 2907 |
| 3 | ESO Observer gain | $p_3$ | 3000 |
| 4 | Roll Proportional-gain | $k_{1p}$ | 90.3979 |
| 5 | Roll Derivative-gain | $k_{2p}$ | 19.6321 |
| 6 | Pitch Proportional -gain | $k_{1t}$ | 79.3794 |
| 7 | Pitch Derivative -gain | $k_{2t}$ | 21.1666 |
| 8 | Yaw Proportional -gain | $k_{1k}$ | 69.8457 |
| 9 | Yaw Derivative -gain | $k_{2k}$ | 16.8096 |
| 10 | Altitude Proportional -gain | $k_{1z}$ | 10.5246 |
| 11 | Altitude Derivative -gain | $k_{2z}$ | 9.5557 |

## B. Results

This section deals with the results obtained from simulation of the quadrotor manipulator system using the ADRC controller. The Altitude reference is set to $z_r = 5m$ while the roll, pitch and yaw angles are set to $\phi_r, \theta_r, \psi_r = 5°$.

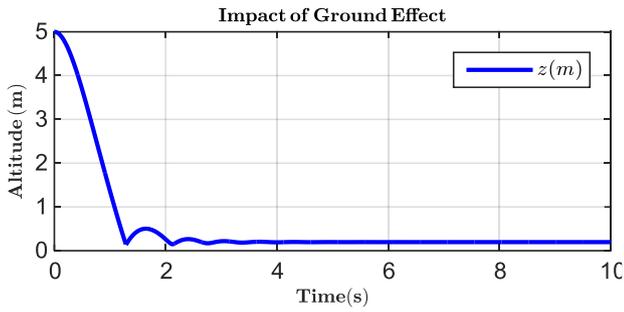

Fig. 5. *Ground Effect on Altitude Dynamics without Controller.*

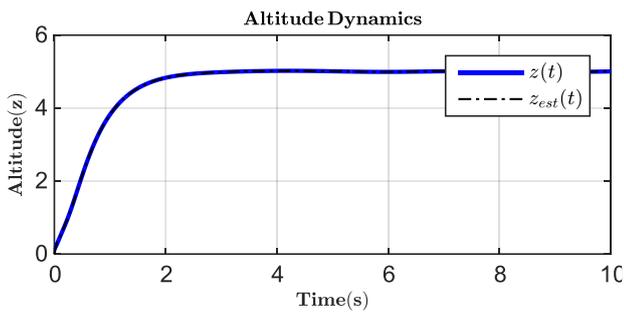

Fig. 6. *True and Estimated Altitude ($z$) of the Quadrotor.*

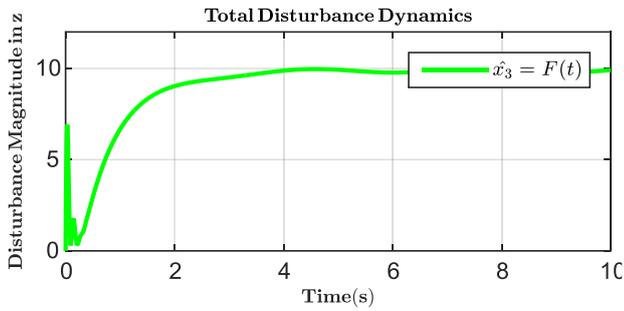

Fig. 7. *Estimated Total Disturbance on Altitude Dynamics*

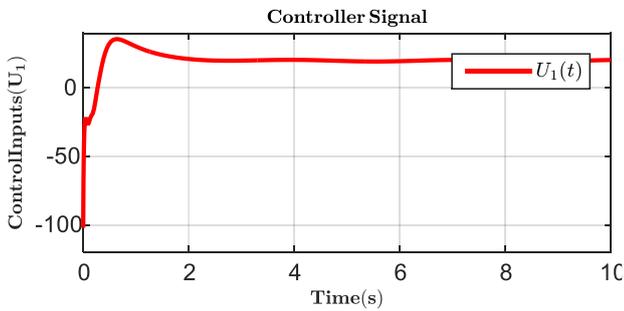

Fig. 8. *Altitude Subsystem Control Input.*

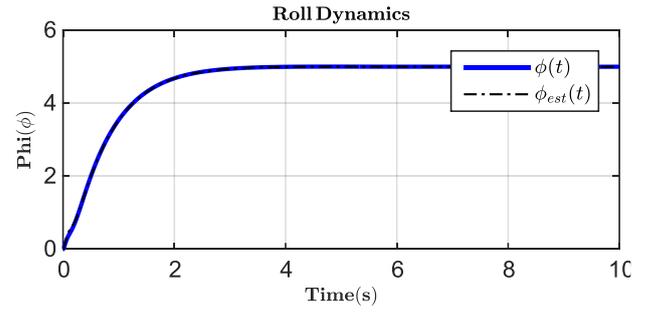

Fig. 9. *True and Estimated Roll Angle ($\phi$) of the Quadrotor*

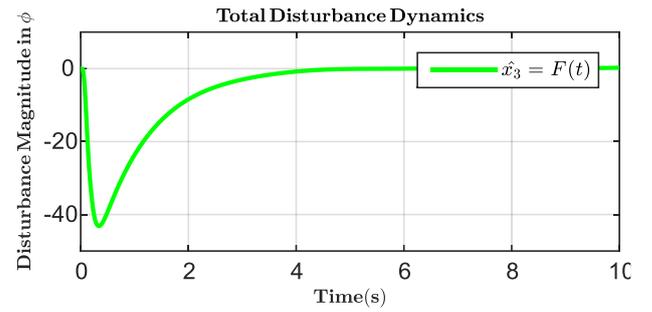

Fig. 10. *Estimated Total Disturbance on Roll Dynamics*

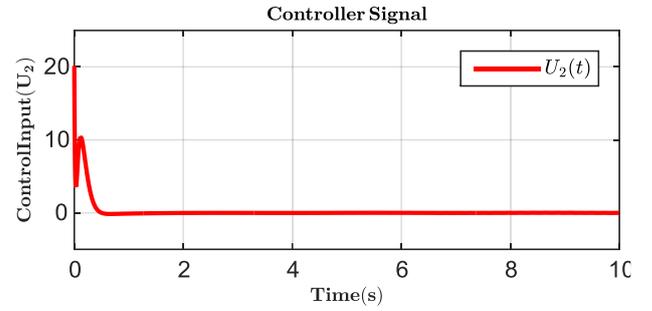

Fig. 11. *Roll Subsystem Control Input.*

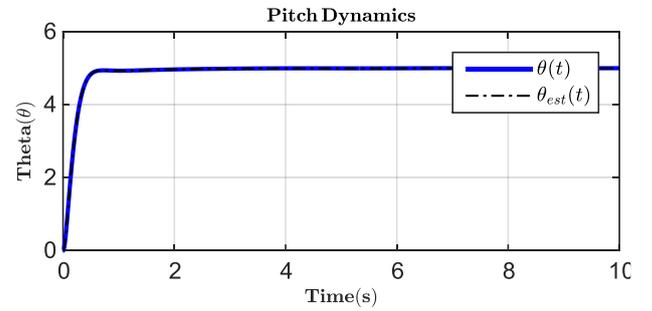

Fig. 12. *True and Estimated Pitch Angle ($\theta$) of the Quadrotor*

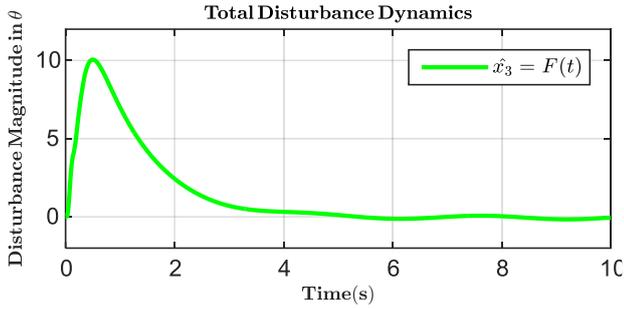

Fig. 13. *Estimated Total Disturbance on Pitch Dynamics*

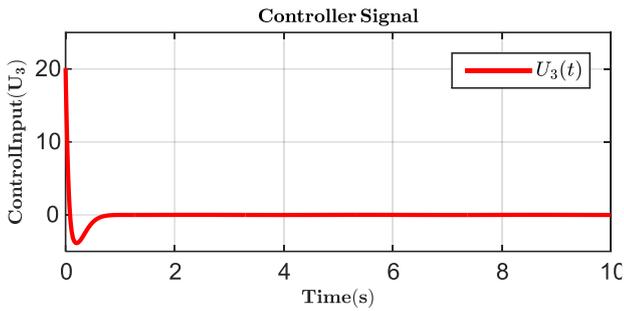

Fig. 14. *Pitch Subsystem Control Input.*

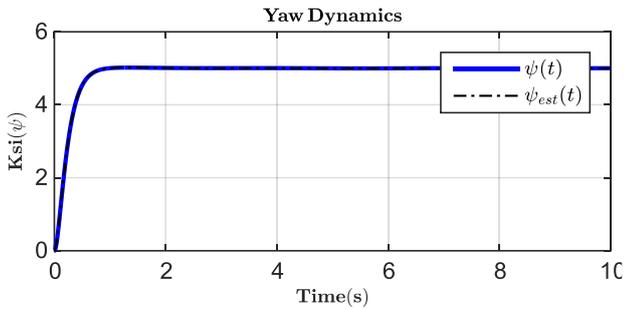

Fig. 15. *True and Estimated Yaw Angle ($\psi$) of the Quadrotor*

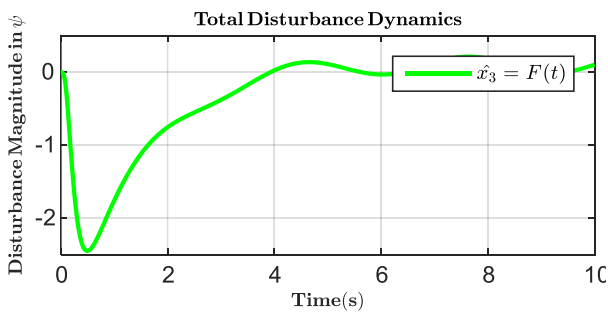

Fig. 16. *Estimated Total Disturbance on Yaw Dynamics*

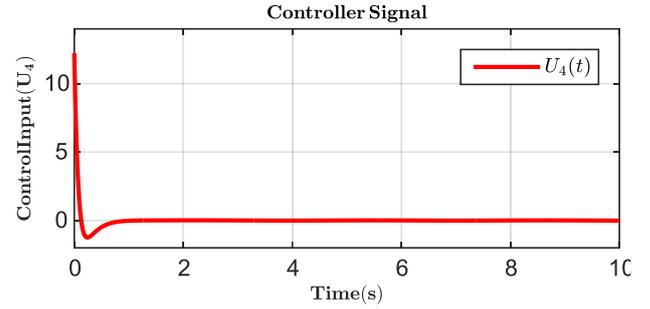

Fig. 17. *Yaw Subsystem Control Input.*

The results are briefly explained as follows: Figure 5 shows the effect of ground wake without appropriate compensation from a controller. The quadrotor bounces at low altitudes and this is undesirable during landing. Figure 6 presents the QM's true and estimated altitude states. In Figures 7, the total estimated disturbance is shown. Figure 7 lumps gravity and other disturbances as a single 'total disturbance'. Figure 8 on the other hand displays the ADRC control inputs to the altitude subsystem. Similar results are plotted in Figures $9 - 17$. The results confirm that the ADRC with proper observer gains is enough to guarantee stability of a Quadrotor Manipulator in the presence of bounded disturbances.

## VI. SUMMARY AND CONCLUSION

In summary, the control of a quadrotor-manipulator was solved using a technique which observes the total disturbances in the system and actively compensates for it. The parameters were tuned to minimize estimation error and constrain the control signal to reasonable limits.

## VII. ACKNOWLEDGEMENTS

I would like to acknowledge the support of Drs. Sami El-Ferik, Samir Al-Amer, and AbdulWahab Al-Saif.## REFERENCES

[1] G. Cai, J. Dias, and L. Seneviratne, "A Survey of Small-Scale Unmanned Aerial Vehicles: Recent Advances and Future Development Trends," *Unmanned Syst.*, vol. 2, no. 2, pp. 175–199, 2014.

[2] M. Orsag, "Mobile Manipulating Unmanned Aerial Vehicle (MM-UAV): Towards Aerial Manipulators," *System*, 2015.

[3] M. Orsag, C. Korpela, M. Pekala, and P. Oh, "Stability